\documentclass[11pt]{article}
\usepackage{hyperref}
\usepackage{authblk}
\usepackage{graphicx}
\usepackage{caption}
\usepackage{amsmath}
\usepackage{amssymb}
\usepackage{mathtools}
\usepackage{multirow}
\usepackage{amsmath}
\usepackage{color}
\usepackage{subcaption}
\usepackage{wasysym}
\usepackage{ulem}

\begin{document}

\title{Medium Assisted Low Energy Nuclear Fusion }

\author{Pankaj Jain$^{1}$\footnote{pkjain@iitk.ac.in}}
\author{Harishyam Kumar$^{2}$\footnote{hari@iitk.ac.in}}
\affil{
	$^1$ Department of Space, Planetary and Astronomical Sciences \& Engineering, Indian Institute of Technology, Kanpur, 208016, India\\
$^2$ Physics Department, Indian Institute of Technology, Kanpur, 208016, India
}

\maketitle

\begin{abstract}
We study the process of nuclear fusion at low energies in a medium 
	using the second
order time dependent perturbation theory. We consider a specific process 
which involves fusion of a low energy proton with a Nickel nucleus. 
The reaction proceeds in two steps or interactions. We refer to
the amplitudes corresponding to these two interactions as 
 the molecular and the nuclear matrix elements. The first amplitude
involves Coulomb interaction with another nucleus in the 
medium while the second corresponds to the nuclear fusion process. 
Due to the presence of high energy intermediate states, the repulsive 
Coulomb barrier may be evaded at this order. However,
it has been shown in earlier papers that contributions from different 
intermediate states cancel one another leading to
 negligible amplitude unless it is assisted by special medium effects.
The medium leads to localization of eigenstates and resultant discretization
of energy eigenvalues which evades the acute cancellation and leads to
observable rate. In the present paper we extend this mechanism
to consider the fusion of a light nucleus with a heavy nucleus.
To be specific we consider the fusion of proton with Nickel to form 
Copper with emission of a photon. The process is assisted by the presence
of an impurity ion in the medium.
\end{abstract}

\section{Introduction}

There exists considerable experimental evidence for nuclear fusion
reactions 
at low energies
 \cite{doi:10.1002/9781118043493.ch41,doi:10.1002/9781118043493.ch42,doi:10.1002/9781118043493.ch43,biberian2020cold,PhysRevC.78.015803,StormsCS2015,McKubre16,Cellani19,Mizuno19,SRINIVASAN2020233}. 
Theoretically, there have been many attempts to explain these processes in terms of electron screening \cite{PhysRevC.101.044609,assenbaum1987effects,ichimaru1993nuclear}, correlated states \cite{articleVy,PhysRevAccelBeams.22.054503}, electroweak interactions \cite{Widom10}, formation of clusters of nuclear particles \cite{SPITALERI2016275}, relativistic electrons in deep orbits \cite{Meulenberg19} and phonon induced reactions \cite{Hagelstein15}.
A critical review of many claims in this field is provided in \cite{Chechin_1994}.

In the present paper we study the possibility that such reactions may 
proceed at second order in time dependent perturbation theory \cite{PhysRevC.99.054620, Jain2020, Jain2021, ramkumar2022, harikumar2023}.  
The first perturbation causes the system to go into a state
which is a linear superposition of all eigenstates of the unperturbed Hamiltonian. Due to the presence of eigenstates of relatively high energy, it is possible that the Coulomb barrier may not be a very serious issue. 
Although the amplitude for such high energy eigenstates is suppressed, 
the suppression may not be as strong as that due to the
 Coulomb barrier. We applied this formalism explicitly to the process involving fusion of proton with deuteron to form helium nucleus with $A=3$ \cite{Jain2020,Jain2021}. The perturbation was assumed to be electromagnetic leading to either emission or absorption of photons. The dominant process was found to be the one in which two photons are spontaneously emitted. 
We need to sum over all intermediate states up to infinite energy. 
We assume that the momentum of the photon emitted at the first (molecular)
vertex is relatively large. 
The amplitude for the process is found to substantial for some fixed values
of intermediate energies, if the relative
proton-deuteron momentum is approximately opposite to the
momentum of the photon emitted at the first vertex.
Here we assume that this momentum is sufficiently large that the 
Coulomb barrier does not lead to a strong suppression. 
Although the amplitude is significant for some range of energies, we find
that as we sum over all intermediate states, the amplitude adds up to very
small values if we assume free space boundary conditions at 
large distances. Hence, the rate is found to be very small in free space.
We obtain contributions from large number of intermediate states
since the eigenstates of the unperturbed Hamiltonian are 
not momentum eigenstates. In particular, due to the Coulomb repulsion, 
the initial state deviates 
considerably from being a momentum eigenstate. 

Although the rate is found to be highly suppressed in free space,
it was argued that in a medium, under special conditions, the rate may be significant and observable \cite{Jain2020,Jain2021}. This was explicitly shown to
work in a simple model \cite{ramkumar2022}.
The basic idea in this paper
is that in a medium the boundary conditions on the wave function at large
distance gets modified, leading to discretization of energy eigenvalues.
Furthermore, in the presence of disorder, the eigenfunctions get localized
\cite{RevModPhys.57.287,doi:10.1142/7663}.  
We point out that phenomenon being discussed here is different from 
Anderson localization \cite{RevModPhys.57.287,doi:10.1142/7663} since we
considering the medium wave function of a proton (or other nuclei) 
and not electron. Furthermore, we are dominantly interested in states
with energy eigenvalues less than the medium potential height. 
If we assume such localized states,
we do not find the acute cancellation of amplitudes that was
found in free space. 
In \cite{ramkumar2022} we used a simple step potential to model the tunneling 
barrier but the mechanism is expected to work also for Coulomb barrier.
The model can directly be applied to fusion of 
light nuclei within a medium composed of heavy nuclei. The energy of 
intermediate states required in this case is of the order of few tens of keV. 
For such energies, it is concievable that the wave functions would be
localized in a medium.

\begin{figure}
     \centering
     \includegraphics[width=0.84\textwidth]{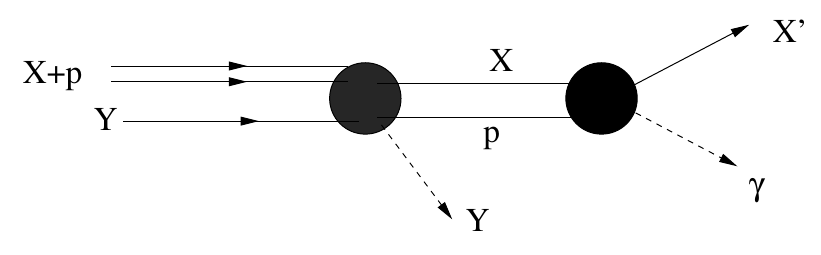}\\
     \caption{Schematic illustration of the reaction described in Eq. \ref{eq:HXfusion}. 
}
\label{fig:process}
\end{figure}

In the present paper we apply this mechanism to the fusion of proton
${}^1$H and a heavy nucleus ${}^A$X of atomic number $Z$ and atomic mass $A$. 
This requires much higher intermediate state energies in comparison to
fusion of two light nuclei and we consider a generalization
of the mechanism
 used in \cite{ramkumar2022}.
At the first vertex,
the process is assisted by an additional particle which we assume
to be a heavy nucleus Y with atomic mass $A_Y$ \cite{PhysRevC.99.054620}. 
Here, we assume $Y$ to be another Nickel nuclei present in the medium. 
At the second vertex proton undergoes fusion with the nucleus $X$ with emission
of a photon.
The basic process can be written as, 
\begin{equation}
 p + {}^{A}{\rm X} 
	+ {\rm Y} \rightarrow {}^{A+1}{\rm X'} + 
	{\rm Y}(\vec k_Y)+ \gamma(\vec k_\gamma) 
    \label{eq:HXfusion}
\end{equation}
where the initial state particles have negligible momentum and
the final state momenta are indicated in this equation. 
The process is illustrated in Fig. \ref{fig:process}.  
Here we are showing
only the nuclear states. In the initial state the proton and the X nucleus 
is assumed to
form a molecular bound state, which is not explicitly indicated in the
above equation. 
The Y atom is present in the 
vicinity and its nucleus interacts with the molecule 
through screened Coulomb interaction. This interaction leads to 
exchange of energy between Y and the molecule and produces a free Y particle
of relatively high momentum \cite{PhysRevC.99.054620}. 
This also breaks up the molecule, producing
two particles with  
high relative momentum. A part of the momentum is transfered to the 
overall center of mass motion. Due to the high
relative momentum, the two particles can now undergo nuclear fusion. 
However we need to coherently add contributions from all intermediate 
states that contribute.

In earlier papers \cite{Jain2020,Jain2021,ramkumar2022}, we had considered emission of a photon at the first vertex.
The main advantage of exchanging momentum with a heavy nucleus Y instead
of a photon is that the nuclear particles can acquire much higher momenta 
in the latter case. Hence, although momentum required for
fusion between two light nuclei at appreciable rates can 
be acquired by emission of a photon, this is not possible in the present
case of fusion between a proton and a heavy nucleus. This 
process, however, suffers from the same problems as those with emission of a photon,
that is, while the amplitude for the process becomes high for some intermediate
states, overall the amplitude tends to cancel as we sum over all states \cite{Jain2020,Jain2021,ramkumar2022}. 
The main point of this paper is the presence of special conditions which
avoid this cancellation.

\section{Proton-X fusion at second order in perturbation theory}

Let us denote the position vectors of the proton and X by $\vec r_1$ and $\vec r_2$. 
 The position vector of the Y particle
is denoted by $\vec r_3$. 
We denote the relative and the CM coordinates of the system of particles 1 and 2 as 
\begin{eqnarray}
	\vec r_{12} &=& \vec r_1 - \vec r_2\nonumber \\
	\vec R_{12} & = & {m_1\vec r_1 + m_2\vec r_2\over m_1+m_2}
\end{eqnarray}
In the initial state we have a molecular bound state along with a free Y particle. 
We can express the wave function as,
\begin{equation}
	\Psi_i(\vec r_1,\vec r_2,\vec r_3) 
=	\psi_i(\vec r_1,\vec r_2) \psi_{Yi}(\vec r_3) 
\end{equation}
The Y particle is assumed to be free and we take its wave function 
to be a plane wave of almost negligible momentum.
The remaining two particles form a bound state.
We assume that the overall center of mass motion acts as a free particle. Hence
the wave function can be expressed as,
\begin{equation}
\psi_i(\vec r_1,\vec r_2) = \psi_{12i} (\vec R_{12})\chi_{12i}(\vec r_{12})
\end{equation}
where the center of mass wave function $\psi_{12i}$ takes the form of a plane wave.  
We assume its momentum to be 
almost zero. 


After Coulomb interaction with the Y particle the molecule breaks up leading to a large relative
momentum between the two particles. 
The center of mass of the proton-X again behaves as  
a free particle. 
The relative coordinate $\vec r_{12}$ dependence of the wave function of this system 
is more complicated since the two particles 
undergo strong Coulomb repulsion along with
the nuclear force at very short distances. 
The intermediate state wave function can be 
expressed as,
\begin{equation}
	\Psi_n(\vec r_1,\vec r_2,\vec r_3) = \psi_{12n}(\vec R_{12})\chi_{12n}(\vec r_{12})\psi_{Yn}(\vec r_3)
\end{equation}
Here the center of mass wave function $\psi_{12n}$ and the $Y$ particle wave function $\psi_{Yn}$ are taken to be plane waves.
The wave function $\chi_{12n}$, which depends on the relative coordinate,
is discussed below.

The Hamiltonian of the system is given by 
\begin{equation}
    H  = H_0 + H_I
\end{equation}
where $H_0$ denote the 
unperturbed Hamiltonian and $H_I$ is the perturbation. 
The unperturbed part contains the kinetic energy terms and screened Coulomb and nuclear potentials
corresponding to the proton-X system. Hence we can express $H_0$ as
\begin{equation}
	H_0 = K_p + K_X + K_Y + V_{12}(\vec r_{12}) 
\end{equation}
where $K_p$, $K_X$ and $K_Y$ represent the kinetic energies of the proton, X and the Y particles
respectively and $V_{12}$ is the potential between the proton and the X nucleus. Here $V_{12}$ 
contains the screened Coulomb potential along with the nuclear potential. 
As discussed later, 
these two particles may also experience the effect of medium at large 
distance which is also part of this potential. 
We assume a shell model nuclear potential, given by,  
\begin{equation}
	V_{nuc} = -{V_0\over 1 + \exp((r-r_{nuc})/r_0)}
	\label{eq:potentialnuc}
\end{equation}
where the parameter $r_{nuc} = 1.25 A^{1/3}$ fm, $r_0= 0.524$ fm and $V_0=36$ MeV. The value of $V_0$ is chosen 
to obtain the expected energy eigenvalue of the nuclear bound state for nucleus under consideration.
Hence the potential takes the form \cite{1968psen.book.....C},
\begin{eqnarray}
	V &=& V_{nuc} \ \ \ \ \ \ \ \ \ \ \ \ \ \ \  \ \ \ \ \ \ \ r\le r_{cut}\nonumber\\
	  &=& {Z_pZ_X\over r} \ \ \ \ \ \ \ \ \ \ \ \ \ \ \ \ \ \ \ \ r> r_{cut}
\end{eqnarray}
where $Z_p$ and $Z_X$ are the atomic numbers of the proton and the X nuclei 
respectively
and $r_{cut}$ is given by \cite{1968psen.book.....C},
\begin{equation}
	r_{cut} = 1.4(A_p^{1/3}+A_X^{1/3})\times 10^{-13} \ \ {\rm cm}
\end{equation}
Here $A_p$ and $A_X$ are the atomic mass numbers of the two particles 
and we have only displayed
the bare Coulomb potential and not the screening part. 

The interaction Hamiltonian can be split into two parts, 
\begin{equation}
	H_I = H_{I1} + H_{I2}
\end{equation}
where $H_{I1}$ contains the Coulomb interaction between proton and Y and 
X and Y while $H_{I2}$  
 is given by \cite{merzbacher1998quantum,sakurai1967advanced}, 
\begin{equation}
	H_{I2}(t) = 
\sum_i{ Z_ie\over c m_i} \vec A(\vec r_i,t)\cdot \vec p_i  
    \label{eq:Hint}
\end{equation}
Here $Z_i$, $m_i$, $\vec r_i$ and $\vec p_i$ are respectively the charge, mass, position vector and momentum vector of the particle $i$. 
The electromagnetic field operator $\vec A(\vec r,t) $ is given by
\begin{equation}
    \vec A(\vec r,t) = {1\over \sqrt{V}} \sum_{\vec k}\sum_\beta c\sqrt{\hbar\over 2\omega} \left[a_{\vec k,\beta}(t) \vec\epsilon_\beta e^{i\vec k\cdot\vec r} +  a^\dagger_{\vec k,\beta}(t) \vec\epsilon^{\,*}_\beta e^{-i\vec k\cdot\vec r}\right]
\end{equation}
We point out that the Coulomb interaction between proton and X must be 
considered as part of the unperturbed Hamiltonian. This is because we are
interested in fusion between these two particles and the full potential
including contributions from
Coulomb repulsion and nuclear attraction must be included in the unperturbed
wave functions. 

The leading order contribution to the process in Eq. \ref{eq:HXfusion} 
is obtained at second order in the time dependent perturbation theory. 
Let the wave vectors of the emitted photon be $\vec k_\gamma$ and the 
frequency be $\omega$. The
transition amplitude at this order 
can be expressed as,
\begin{equation}
	\langle f|T(t_0,t)|i\rangle = \left( -{i\over \hbar}\right)^2\sum_n \int_{t_0}^t dt' e^{i(E_f-E_n)t'/\hbar} \langle f|H_I(t')|n\rangle \int_{t_0}^{t'} dt'' e^{i(E_{n}-E_i)t''/\hbar} \langle n|H_I(t'')|i\rangle
    \label{eq:TransitionII} \ .
\end{equation}
where the sum is over the intermediate proton-X states.
At the first vertex, the Y particle interacts with the proton and the X nucleus through the
screened Coulomb potential, i.e. the interaction Hamiltonian $H_{I1}$
\cite{PhysRevC.99.054620}.
We refer to the corresponding amplitude as the
molecular matrix element. At the second vertex the proton and X undergo 
nuclear fusion while emitting a photon through the interaction 
Hamiltonian $H_{I2}$. The corresponding amplitude is 
called the nuclear matrix element.

\subsection{Molecular Matrix Element}
Let us first consider the transition from initial to intermediate state 
through the Coulomb interaction. The matrix element can be written as
\begin{equation}
	\langle n|H_{I1}|i\rangle =  \int d^3r_1d^3r_2d^3r_3
	\Psi_n^*\left[ V_s(p,Y) + V_s(X,Y)\right]\Psi_i
\label{eq:inttpp}
\end{equation}
where $V_s(a,b)$ is the screened Coulomb potential
\cite{PhysRevC.99.054620},
\begin{equation}
	V_s(a,b)
	= {Z_a Z_b e^2\over 2\pi^2} \int d^3q {e^{i\vec q\cdot (\vec r_a-\vec r_b)}\over q^2 + q^2_s} 
\end{equation}
Here $q_s$ is the contribution due to screening. 
As we will see the contribution from the second term in Eq. \ref{eq:inttpp} will
be relatively small and we focus on the first term. We change variables to
$\vec r_{12}$, $\vec R_{12}$ and $\vec r_3$ and set 
 $\psi_{Yn} = e^{i\vec k_3\cdot \vec r_r}/\sqrt{V}$ and $\psi_{Yi}\approx
1/\sqrt{V}$. Here we have assumed negligible momentum for Y in the initial
state. After integrating over
$\vec r_3$ we obtain,
\begin{equation}
	\langle n|H_{I1}|i\rangle =  {4\pi Z_pZ_Ye^2\over V}
	\int d^3r_{12}d^3R_{12}d^3q
	\psi_{12n}^*\chi_{12n}^*  e^{i\vec q\cdot r_1}{\delta^3(\vec k_3+\vec q)
	\over q^2+q_s^2} \psi_{12i}\chi_{12i} + ...
\label{eq:inttpp1}
\end{equation}
Integrating over $\vec q$ we obtain
\begin{equation}
	\langle n|H_{I1}|i\rangle =  {4\pi Z_pZ_Ye^2\over V}
	\int d^3r_{12}d^3R_{12}
\psi_{12n}^*\chi_{12n}^*  e^{-i\vec k_3\cdot \vec r_1}{1
	\over k_3^2} \psi_{12i}\chi_{12i} + ...
\label{eq:inttpp2}
\end{equation}
Here we have dropped $q^2_s$ since we shall set $k^2_3>>q^2_s$.  
We next replace $\vec r_1 = \vec R_{12} + \vec r_{12} m_2/(m_1+m_2)$ 
and take the initial and intermediate center of mass wave functions to be 
plane waves. The intermediate center of mass momentum is taken to be 
$\vec K_{12}$ while the initial momentum is assumed to be negligible. 
Integration over $d^3R_{12}$ leads to a delta function $\delta^3(\vec K_{12}
+\vec k_3)$ and we focus on the integral in terms of relative variables.
We obtain
\begin{equation}
\langle n|H_{I1}|i\rangle =  {4\pi Z_pZ_Ye^2\over Vk_3^2}
I_i + ...
\label{eq:inttpp3}
\end{equation}
where 
\begin{equation}
	I_i=
	\int d^3r_{12}
	\chi_{12n}^*  e^{-i\vec k_3\cdot \vec r_{12}m_2/(m_1+m_2)}
	\chi_{12i}^{ } 
\end{equation}
In the exponent the ratio of masses is approximately unity with $m_2>>m_1$
($m_X>>m_p$). We will assume that the magnitude of $k_3$ is very large.

We next discuss the wave function $\chi_{12i}$ and $\chi_{12n}$ which 
depend on the relative coordinate $\vec r_{12}$. At short distance both
of these are determined by the spherically symmetry nuclear potential along
with the Coulomb repulsion. However their large distance behaviour is
determined by the medium effects. The important point of this paper is 
that this large distance behaviour has very significant effect on the
nuclear reaction. Hence by suitably choosing or tuning the medium we 
can change the nuclear reaction rate by orders of magnitude. 
We point out that in free space, assuming spherical symmetry,
the reaction rate is negligible. 

The initial state wave function $\chi_{12i}$ is an eigenstate of
the unperturbed Hamiltonian $H_0$, corresponding to 
a very low energy eigenvalue. 
We assume this to be bound state wave function.  
The proton experiences the spherically symmetric repulsive Coulomb potential of the nucleus $X$
at short distances and  
hence the wave function decays very strongly at distances $r_{12}<1$. At intermediate distances,
i.e. distances scales of a few Bohr radius, it experiences 
the attractive molecular potential. At larger distances it experiences the
repulsive potential due to all the other ions present in the medium. 
The medium potential is likely to be complicated and is not expected to
show any simple symmetry, especially in the presence of disorder. 
Here we do not attempt 
a detailed computation of the wave function and make simple
assumptions about its nature. We assume that the
wave function has roughly the same behaviour in all directions and hence 
can be approximated to be a function of only the radial distance $r_{12}$.
Furthermore, we assume the potential to be 
approximately constant over the range 
$\Delta \le r_{12}\le r_u$ and strongly repulsive outside this range. 
The wave function can, therefore, be expressed as
\begin{equation}
	\chi_{12i} = NY_0^0 {\sin[(r_{12}-\Delta)/r_0]\over r_{12}}
\label{eq:chii}
\end{equation}
for $\Delta \le r_{12} \le \pi r_0+\Delta$ and zero outside this range. 
The wave function can be smoothed at the two ends, as applicable for a finite
potential barrier. However, this does not affect our
results significantly. Here $N$ is the normalization, the parameter
$\Delta$ is taken to be about 1 atomic unit and $r_0$ is of the order of
the typical molecular distances.

In order to compute the total amplitude we need to sum over all the
intermediate state wave functions $\chi_{12n}$ that can contribute. 
These are also eigenstates of the unperturbed Hamiltonian $H_0$. 
The dominant contribution is
obtained from eigenstates of relatively large energy eigenvalue.
At small distance these are solutions 
to the spherically symmetric nuclear potential including the Coulomb
barrier.  
At large distances their behaviour is also determined by the molecular potential and the medium
properties. 
Due to their high energy, these states may extend over the entire medium
and, hence, the large scale structure of the medium has to be considered,
which does not display spherical symmetry.  
For example, in a periodic crystalline lattice, all eigenstates are of the 
form of Bloch functions, $\psi(\vec r) = u(\vec r) e^{i\vec k\cdot \vec r}$
where the function $u(\vec r)$ also displays the periodicity of the 
lattice. Here we do not go into a detailed solution of the Schrodinger
equation in a medium and make some physically motivated assumptions. 
We assume that the medium is not periodic and displays disorder. In such a
case, the eigenfunctions up to some maximum energy eigenvalue must be 
localized 
\cite{RevModPhys.57.287,doi:10.1142/7663}.  
Here we are interested in eigenstates with very high energy, but
whose wave vectors $\vec k_{12}$ are large only in one direction, i.e.
$\vec k_{12} \approx -\vec k_3$.
We choose the z-axis along $\vec k_3$. The $z$ dependence of
the wave function can be approximated as $e^{ik_zz}$. This is not significantly affected
by the medium, which may lead to a slow modulation of the plane wave behaviour
which is being ignored here.
Due to the large value of $|k_z|$ we assume that it takes almost continuous 
range of values.  
However, in the transverse
direction the behaviour of the wave function is determined by the transverse
wave numbers $|k_x|$ and $|k_y|$ which are very small compared to $|k_z|$. 
Hence it is reasonable to assume that in these directions
the wave function may be significantly modulated by the behaviour of the 
medium potential, leading to localization of wave function in these
directions and discretization of $k_x$ and $k_y$. For simplicity, we assume  
that the wave function displays approximate
cylindrical symmetry and is a function only of $\rho=r\sin\theta$. 
We denote it by $f(\rho)$ and  
 write the full wave function as,
\begin{equation}
	\chi_{12n}(\vec r_{12})= {1\over \sqrt{L}} e^{ik_zz}
	f(\rho) 
	\label{eq:wavechin}
\end{equation}
where $k_{12}^2 = k_z^2+k_\rho^2$ and $k_\rho<< |k_z|$.
Excluding very small values of $\rho$, comparable to nuclear distances, 
the wavefunction $f(\rho) \approx N_n(k_\rho) J_0(k_\rho\rho)$, where
$J_0$ is the Bessel function and $N_n$ is the normalization factor. 

As discussed above, the function $f(\rho)$ is localized in space. We assume
that it is zero for $\rho > \rho_0$. For simplicity, we take the upper length
scale $\rho_0$ to be equal to $r_u = \pi r_0+\Delta$ 
which is same as that in the case of 
$\chi_{12i}$. 
Such a boundary condition is applicable for an infinite potential. For the
case of a finite potential, we expect a smooth transition. However, this
is not expected to significantly affect our results.
Furthermore, by explicit calculations, we find that the amplitude decreases 
relatively slowly as we increase the value of $\rho_0$ and, hence, the chosen
value gives a reasonable description of the physical process under 
consideration.

At small distances, of order nuclear scale, we need to match this 
solution to the nuclear tunneling wave function. This is obtained by
solving the Schrodinger equation in the spherically symmetric nuclear 
potential including the Coulomb repulsion. The dominant contribution is
obtained from the $l=0$ wave function, which 
depends on $k=\sqrt{k_z^2+k_\rho^2}$. For the values
of $E_n$ of interest, with $|k_z| >> k_\rho$, it only shows a
very mild dependence on $k_\rho$. 
Hence the nuclear matrix element cannot depend significantly
on the value of $k_\rho$ which is much smaller than $k_z$. 
Therefore, it is reasonable to   
assume that the projection of $f(\rho)$ on the spherically symmetric $l=0$
short distance wave function
is approximately constant, independent of $k_\rho$ for small $k_\rho$. 
We approximate the constant value of this wave function
for small $\rho$
to be $N_n(k_{\rho 1})$, i.e. the normalization
corresponding to the first zero of the Bessel function.
We clarify that the actual value of the function $f(\rho)$ at very small $\rho$
may be somewhat 
different from the value assumed here, however, the essential point is that
it cannot depend significantly on $k_\rho$. 

The form of the wave function given in Eq. \ref{eq:wavechin} follows
by solving the Schrodinger equation using an approximation similar 
to the Born-Oppenheimer 
approximation. We note that wave function varies very rapidly along the
z-direction and slowly along the transverse direction. Hence, we first
solve the Schrodinger equation to determine the $z$ dependence 
for a fixed $\rho$. The solution obtained leads to an effective potential
for the $\rho$ dependence and we can obtain the $\rho$ dependence by
solving the resulting equation. This is illustrated in Appendix A.
The essential assumption in obtaining Eq. \ref{eq:wavechin} is that the
$z$ dependence can be taken to be a plane 
wave to a good approximation. This is reasonable since the corresponding wave 
number $k_z$ is very large. However, as we shall see later, dominant
contribution is obtained from small values of the transverse wave number 
$k_\rho$. Hence, the corresponding wave function may be strongly affected
by the medium potential and is expected to decay
rapidly at large $\rho$ leading to localization in the transverse
plane and discretized values of $k_\rho$.

As mentioned above, 
we also need to match the large and short distance wave functions.
The function $f(\rho)$ goes to a constant at small distance and
we use the standard expansion for the plane wave,
\begin{equation}
	e^{i\vec k\cdot \vec r} = 4\pi \sum_{l=0}^{\infty}
	\sum_{m=-l}^{l} i^l j_l(kr)\ Y_l^{m}(\hat r)
	Y_l^{m*}(\hat k)
	\label{eq:PlaneWave}
\end{equation}
For energies under consideration the $l=0$ part will dominate and the high 
angular momentum contributions will be strongly suppressed due to  
Coulomb repulsion. We solve the
Schrodinger equation for small $r_{12}$ in the presence of the nuclear
and the Coulomb potential for $l=0$. The difference between this and
the $l=0$ component of Eq. \ref{eq:PlaneWave} is a scattered wave which
will decay at large $r$ and hence can be neglected. 
The wave function may deviate from the assumed form at intermediate
distances. However these play negligible role in the determination of
the amplitude.

Before ending this section we explain why 
the contribution due the second term on the right hand
side in Eq. \ref{eq:inttpp} is relatively small. This is because
the exponent in Eq. \ref{eq:inttpp3} in this case depends on the 
mass ratio $m_1/(m_1+m_2)$ which is approximately $m_p/m_X<<1$. 
In contrast, in the first term the mass ratio is approximately unity 
and dominant contribution is obtained for $k_{12}$ of order $k_3$. 
Hence, 
the dominant contribution from second term
will be obtained for $k_{12}$ much smaller than $k_3$ and 
will be suppressed.

\subsection{Nuclear Matrix Element}

We next consider the nuclear matrix element $\langle f|H_I(t')|n\rangle$. 
In this case, the intermediate state proton undergoes 
fusion with the heavy nucleus with photon emission. The process is complicated since as the proton interacts with the nucleus, it can interact with all the nucleons, leading to a change in the multiparticle wave function. 
Here we assume that we can treat the nucleus $X$ as a 
single particle of charge $Z_X$. 
 We denote the emitted photon momentum by $\vec k_{\gamma}$ and energy
 by $E_\gamma$. 
Let the angular coordinates of this photon momentum be $(\theta_{\gamma}, 
\phi_{\gamma})$, i.e.,  
\begin{equation}
	\vec k_{\gamma } = k_{\gamma } \left[\cos\theta_{\gamma }\, \hat z 
	+\sin\theta_{\gamma }(\cos\phi_{\gamma }
\,	\hat x + \sin\phi_{\gamma }\, \hat y)\right] 
\end{equation}
The two polarization vectors of this photon can be expressed as
\begin{eqnarray}
	\vec\epsilon_{a} &=& 
	 -\sin\theta_{\gamma }\, \hat z +\cos\theta_{\gamma }
	 (\cos\phi_{\gamma }
	\, \hat x + \sin\phi_{\gamma }\, \hat y) \nonumber\\
	\vec\epsilon_{b} &=& -\sin\phi_{\gamma }\,  \hat x + 
	\cos\phi_{\gamma }\, \hat y
\end{eqnarray}

In order to proceed further we need to specify the final state eigenfunction.
We take this state to be $l=1$, $j=3/2$ shell model 
state with one unpaired proton in the
outer most shell. We shall consider the state for which 
proton has spin up and set $j_z=3/2$. 
We point out that ${}^{62}$Ni (spin 0)
satisfies
our requirements. With addition of one more proton, this makes a transition
to ${}^{63}$Cu.
Using the nuclear potential (Eq. \ref{eq:potentialnuc}), we find that the
energy eigenvalue of this state is 9.1 MeV in good agreement with the observed value. 

We perform the calculation by specializing to a particular polarization
vector $\vec \epsilon_b$ of the emitted photon. 
Let $\vec R'$ be the final state center of mass coordinate of the 
proton-X system and ${\vec r}\,'$ be the corresponding relative coordinate. 
The final state wave function
is expressed as
\begin{equation}
	\Psi_f(\vec R',\vec r\,') = \psi_f(\vec R')\chi_f(r')
\end{equation}
The center of mass wave function $\psi_f$ is assumed to be a plane wave.
The wave function $\chi_f$ is taken to be the $l=1$ nuclear shell model
bound state wave function and we express it as
\begin{equation}
	\chi_f(\vec r') = N_fY_1^1{u_f(r')\over r'} 
\end{equation}
Furthermore, as argued above
at short nuclear distance only the $l=0$ part of the intermediate
wave function $\chi_{12n}$ is expected to dominate. 
Hence we set it equal to
$\chi_{12n}= NY^0_0u_n(r')/r'$. 
The matrix element can be expressed as,
\begin{equation}
	\langle f|H_I(t')|n\rangle = -ie\left({Z_Xm_p-m_X\over m_p+m_X}\right) 
	\sqrt{1\over
	8\pi V\omega\hbar}\, e^{iE_{\gamma}t'/\bar h} (E_f-E_{n})
	I_\Omega I_RI_f
\end{equation}
where $E_n$ and $E_f$ are the intermediate and final state energy eigenvalues,
\begin{equation}
	I'_{\Omega} = i\, \sqrt{2\pi\over 3}e^{-i\phi_{\gamma 2}} \,,
	\end{equation}
\begin{equation}
I_R	= \int d^3R' \psi^*_f(\vec R')
	e^{-i\vec k_\gamma\cdot \vec R'}  \psi_{n12}(\vec R') 
\end{equation}
and
\begin{equation}
I_f	= \int dr' u^*_f(r')
	r'  u_{nX}(r') \,,
\end{equation}
The center of mass integral simply imposes overall momentum conservation
$\vec K_f+\vec k_\gamma +\vec k_3=0$, where $\vec K_f$ is the center of mass
momentum of the final state nucleus,
and we focus here on
 the relative coordinate integral $I_f$. 
In this amplitude we have made the standard approximation of
neglecting the momentum of the photon since this
integral gets dominant contribution from small $r'$.
We compute it for a range of 
 energy eigenvalues $E_n$. For small $E_n$ this integral is very small due 
 to strong Coulomb barrier but increase rapidly with increase in $E_n$. 
For the range of $E_n$ which correspond to values of $k_{12}$ close to 
$k_3$ the integral is relative large and does not show a strong dependence
on $E_n$.

\section{Reaction Rate}
Using the molecular and nuclear matrix elements we can compute 
the transition matrix element given in Eq. \ref{eq:TransitionII}. 
The corresponding reaction rate is given by,
\begin{equation}
{dP\over dt} = {1\over \Delta T}\int {Vd^3k_3\over (2\pi)^3} dE_{\gamma} \rho_{\gamma} 
|\langle f|T(t_0,t)|i\rangle|^2
\end{equation}
where $\Delta T$ is the total time and $\rho_{\gamma}$ is the corresponding 
number density of photon states, given by,
\begin{equation}
    \rho_{\gamma } = {V\omega^{ 2}\over (2\pi)^3} {d\Omega\over \hbar c^3}
    \label{eq:photonDOS}
\end{equation}
The time integral in the transition matrix element is proportional to $\Delta T\delta(E_f-E_i+E_{\gamma} + E_{3})$.   
The reaction rate can now be expressed as
\begin{equation}
{dP\over dt} = {Z_3^2\alpha^3 \over 6\hbar V}
\left({Z_Xm_p-m_X\over m_p+m_X}\right)^2 
	\int { d^3k_3\over k_3^4}E_\gamma  
	\ \Bigg| \int {Ldk_{z}\over 2\pi}\sum_{k_\rho} 
	I_f  
	{(E_f-E_n)\over E_n+E_3-E_i}I_i\Bigg|^2
	\label{eq:rate3}
\end{equation} 
The factor $1/V$ corresponds to the normalization of the initial
wave function of $Y$ particle. We need to replace this with the actual
number density $n_Y$ of this particle \cite{PhysRevC.99.054620}.
The rate depends on the two integral $I_i$ and $I_f$ which get
dominant contributions from molecular and nuclear distances respectively. 
The integral $I_f$ depends on the Coulomb repulsion, which for large
intermediate energies $E_n$, is not very prohibitive.  
The most important part of the above formula is contained in $I_i$. 
This is counterintuitive since this integral is controlled entirely
by Physics at distances larger than an atomic unit. Yet, as seen earlier
in \cite{ramkumar2022}, 
this dictates the entire process.

The integral $I_i$ can be expressed as
\begin{equation}
	I_i = {2\pi N_n(k_\rho)\over \sqrt{L}}\int dr_{12} r^2_{12}d\cos\theta
	e^{-i(k_z+\bar k_3)r\cos\theta}J_0(k_\rho r\sin\theta)	
	\chi_{12i}
\end{equation}
where $\bar k_3= k_3 m_2/(m_1+m_2)$. 
 Notice that we have set $\chi_{12i}$ to be zero for 
length scales smaller than approximately 1 atomic unit 
due to the strong Coulomb 
repulsion in this region. Since this wave function is strongly cutoff at small distances, this integral does not lead to a delta function in momentum. 
This
means that we do not have momentum conservation at the first vertex and
is nonzero for a range of values of $\vec k_{12}$. 
However, as we shall see, the range is relatively narrow.
Over this narrow range the integral $I_f$ is almost independent of $k_{12}$
and we can get some idea about the total rate by computing the integral 
over $k_{z}$ and sum over $k_\rho$, setting $I_f$ to unity.

\begin{figure}
     \centering
     \includegraphics[width=0.84\textwidth]{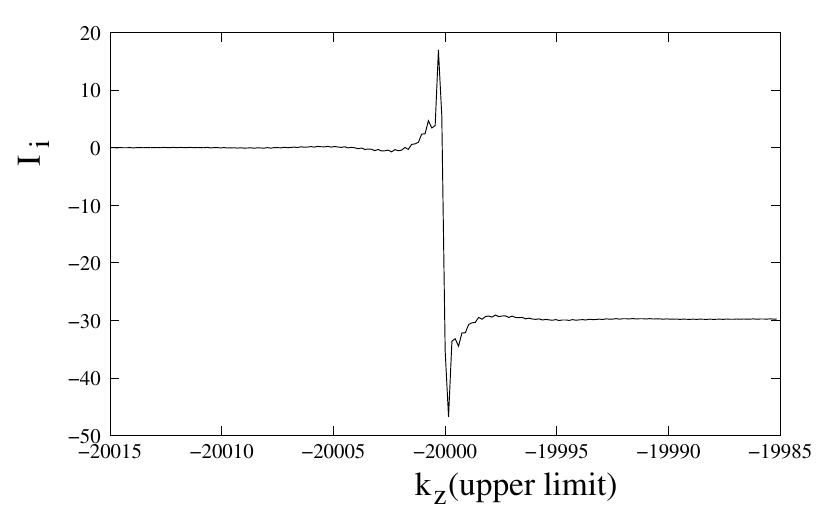}\\
	\caption{The integral in Eq. \ref{eq:Ik} as a function of the upper
	limit on $k_z$ after summing over 
	$k_\rho$. }. 
\label{fig:kzdep}
\end{figure}

\begin{figure}
     \centering
     \includegraphics[width=0.84\textwidth]{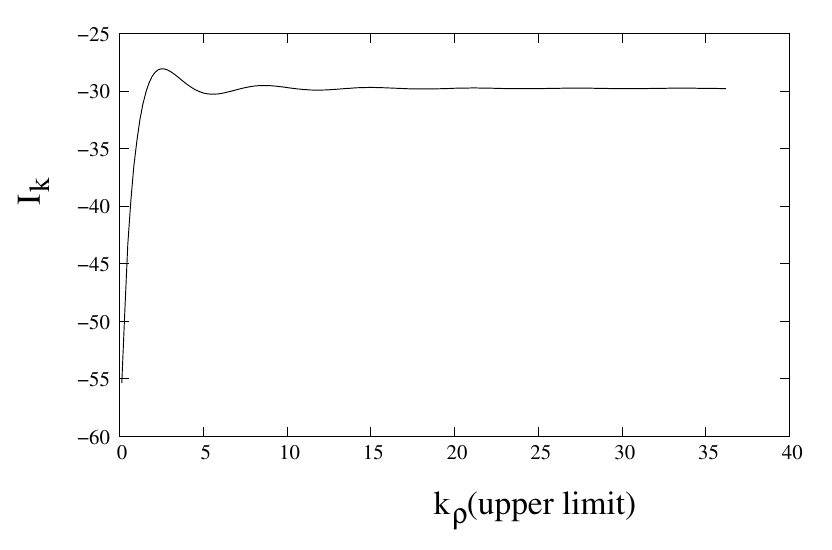}\\
	\caption{The integral in Eq. \ref{eq:Ik} as a function of the
	upper limit on $k_\rho$ after integrating over 
	$k_z$. }. 
\label{fig:kpdep}
\end{figure}

 We consider the integral
\begin{equation}
	I_k = \sqrt{L}\int dk_{z}\sum_{k_\rho}
	{(E_f-E_n)\over E_n+E_3-E_i} I_i 
	\label{eq:Ik}
\end{equation}
which essentially controls the rate of the process. 
The imaginary part of the integral is found to be negligible and we focus
on the real part. 
 In Fig. \ref{fig:kzdep} we plot $I_k$  
 as a function of the upper cutoff on $k_{z}$, after
summing over $k_{\rho}$.  
 In Fig. \ref{fig:kpdep} we plot $I_k$  
 as a function of the upper cutoff on $k_{\rho}$, after
summing over $k_{z}$. We clearly see that the integral saturates to 
a finite, nonzero value and hence we expect the rate to be appreciable.
We point out that the nuclear matrix element shows very slow dependence on 
$k_{12} = \sqrt{k_z^2+k_\rho^2}$ over the small range in which $I_k$ shows
significant variation. Hence we may assume the nuclear matrix element to
be approximately constant in this range.

\begin{figure}
     \centering
     \includegraphics[width=0.84\textwidth]{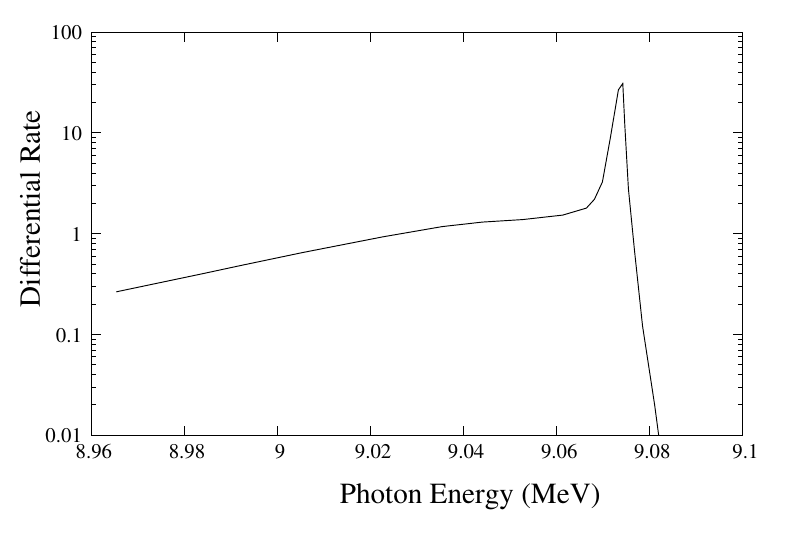}\\
	\caption{The differential rate, $10^{38}\times d^2P/dtdk_3$, in 
	atomic units as a function of the photon energy. 
	 }. 
\label{fig:rate}
\end{figure}

To compute the final rate we set the number 
density of $Y$ particles to be equal 
to $2.65\times 10^{20}$ cm$^{-3}$, same as that in Ref. \cite{PhysRevC.99.054620}. The differential rate, $d^2P/dtdk_3$, as a function of the emitted photon 
energy is shown in Fig. \ref{fig:rate}. We see that the rate increases 
slowly with increase in photon energy, followed by a peak a little above
9.07 MeV and a sharp decline at higher photon energies. The
peak can be attributed to the presence of a resonance at this energy. 
The sharp decline arises since the energy of the $Y$ particle becomes
relatively small leading to a small momentum transfer at the first 
vertex. Hence, the Coulomb barrier strongly suppresses the rate at this region. 
Besides this, we also start hitting the kinematic limit imposed by 
$Q$ value of the process. The total rate for the process turns out to
be about $2\times 10^{-17}$ per second, which is clearly observable in 
standard cold fusion experiments. This implies that if we have $10^{17}$ 
pairs of $H-X$ molecules in the experimental set up we expect roughly two
events per second. 

Experimentally the nuclear transmutations in Nickel using electrolysis has
been explored in \cite{Miley96,Rajeev17,AKumar2022}. Transmutation to Copper
has also been reported in these papers. However, so far there does not exist 
any evidence of gamma ray emission in such experiments. From our analysis we
see that the dominant emission is likely to happen at relatively large values
of photon energies. It is not clear if such an energy regime has so far been
explored. Our results provide a clear signal which can be explored in future 
experiments. We also add that we have only considered one process. Many other
final states are possible which we postpone to future research.

\section{Conclusions}
In this paper we have computed the rate for the nuclear reaction given in Eq. \ref{eq:HXfusion} at second order in time dependent perturbation theory. The initial state consists of a proton and a heavy nucleus ${}^A$X at very low energies, of order eV or less. The reaction proceeds by Coulomb interaction with a 
third particle which leads to the formation of an intermediate state with large
relative momentum between the proton and $X$ particle. Due to the large momentum the Coulomb repulsion between the two is not very prohibitive and the
amplitude for fusion process is not suppressed. We need to sum over all the 
intermediate states. Assuming standard free space boundary conditions on the
corresponding wave functions at large distance, we find that the sum over 
all states involves a very delicate cancellation and leads to a very small
rate. 
We argue that in medium the boundary conditions can be very different.
The intermediate state has relatively high momentum. We assume that in the
direction of this momentum the wave function behaves as a plane wave. 
However, in transverse directions, dominant contribution is obtained from
relatively small momentum components. Hence, the wave function in these
directions would be relatively localized due to medium effects. Based
on these arguments, we assume a cylindrically symmetric wave function of
the form given in Eq. \ref{eq:wavechin}. 

The nuclear process involves fusion of proton with Nickel to produce Copper
with emission of a photon. We find the rate to be sufficiently large to
be observable in laboratory. The emitted photon energy is predicted to
be relatively high of the order of 9 MeV. To the best of our knowledge, 
emission of such high energy photons have not been probed in these processes
so far. Hence, it will be very interesting to experimentally test this 
prediction corresponding to the rates provided in this paper.

Our main goal in this paper is to determine
if the phenomenon of LENR is possible within
the framework of the second order perturbation theory. Hence, in our
analysis we restricted ourselves to physically motivated wave functions.
A more detailed description would require modelling of medium potential 
in a disordered system, which we have postponed to future research. 
 Our results clearly show that LENR is indeed possible. Many other nuclear 
processes, besides the
one considered in this paper, are possible. Some of these may have higher
rates in comparison to the one considered in this paper and should be
investigated in future research.

 \bigskip
 \noindent
 {\bf Acknowledgements:} We are grateful to K. P. Rajeev and K. Ramkumar for 
 useful discussions.

\section{Appendix A}
In this Appendix we illustrate the basic idea behind the construction of the
intermediate state wave function $\chi_{12n}$. The potential is complicated 
since it involves contributions from a large number of ions in a disordered
medium. 
Let us first consider the initial state wave function $\chi_{12i}$. For 
simplicity, we may assume that over a small distance of a few Bohr radii, 
the potential may be approximated to be spherically symmetric.
As mentioned in text, the potential may be taken to be constant over the
range $\Delta\le r_{12}\le r_u$ and strongly repulsive outside this range.
For simplicity we set the constant value of the potential to be zero.
This leads to the initial state wave function given in Eq. \ref{eq:chii}.
We point out that potential is being approximated as spherically symmetric
only in the small region. For the initial state, this is the only region
that is relevant. However, the intermediate state spreads over the entire
medium where spherical symmetry is not valid. Hence, it need not be a 
solution to a spherically symmetric potential.

To determine the intermediate state wave function we note that 
dominant contribution is obtained from states
of the form $e^{ik_zz}$ with $k_z$ much larger than the inverse of Bohr radius. 
Hence, these states have
very rapid variation along the $z$ direction and relatively slow variation
in the transverse direction. 
This suggests the use of cylindrical coordinates and we assume that the 
wave function takes the form, 
\begin{equation}
	\psi(\vec r)= g(\rho,z)	f(\rho) 
	\label{eq:wavechin1}
\end{equation}
where, for simplicity, we have assumed that there is no dependence on the
azimuthal angle $\phi$.
 We use the 
Born-Oppenheimer procedure and first determine $g(\rho,z)$ for fixed
$\rho$. The Schrodinger equation leads to,
\begin{equation}
-{ \hbar^2\over 2m}{\partial^2\over \partial z^2}g(\rho,z)   
	+V(\rho,z) g(\rho,z) = E_z(\rho)g(\rho,z)
\end{equation}
Having solved this equation for all values of $\rho$ we substitute the
solution in the full Schrodinger equation. The dominant terms in the 
resulting equation are given by
\begin{equation}
-{\hbar^2\over 2m}{1\over\rho}{\partial \over \partial \rho}\left(\rho{\partial
	f(\rho)\over \partial\rho}\right) + E_z(\rho)f(\rho) = E f(\rho) 
\end{equation}
The term $E_z(\rho)\approx E$ leads to an effective potential for the $\rho$
dependence of the wave function. 
In obtaining this equation we have dropped terms which involve  derivatives
of $g$ with respect to $\rho$. This is reasonable since the $\rho$ dependence
of $g$ is expected to be very small.
Given a potential $V(\rho,z)$ one can solve the Schrodinger equation using 
this procedure and determine the wave function.

In the current paper, we do not directly solve the Schrodinger equation and use 
a physically motivated wave function. We argue that $g(\rho,z)$ is 
dominantly of the form $e^{ik_zz}$ with constant $k_z$. This is because
the potential is negligible compared to the energy eigenvalue. There is a
slow modulation with $\rho$ which is being neglected in our analysis.
Furthermore, we require the detailed form of the effective potential $E_z(\rho)$
in order to determine the $f(\rho)$. Here we have simply assumed that it is
relatively small over intermediate distances and large at $\rho$
of order $r_u$. These assumptions lead to the wave function given in  
Eq. \ref{eq:wavechin}.

\bibliographystyle{ieeetr}
\bibliography{nuclear}
\end{document}